\RequirePackage{fix-cm}
\documentclass[runningheads]{llncs}       % onecolumn (second format)
\usepackage{cite}
\usepackage{amsmath,amssymb,amsfonts}
\usepackage{algorithmic}
\usepackage{graphicx}
\usepackage{textcomp}
\usepackage[table]{xcolor}
\usepackage{array}
\usepackage{float}
\usepackage{booktabs}
\usepackage{threeparttable}
\usepackage{subcaption}
\usepackage{array}
\usepackage{xurl}
\usepackage{multirow}
\usepackage{hyperref}
\begin{document}

\title{Generative AI Misuse Potential in Cyber Security Education: A Case Study of a UK Degree Program%\thanks{Grants or other notes
}

\titlerunning{Generative AI Misuse Potential in Cyber Security Education}        % if too long for running head

\author{Carlton Shepherd}
\institute{
\begin{tabular}{cc}
Department of Computer Science & School of Computing \\
Durham University & Newcastle University \\
Durham, United Kingdom & Newcastle, United Kingdom
\end{tabular}\\
\email{carlton@linux.com}
}


%\institute{C.\ Shepherd \at
 %             School of Computing\\
 %             Newcastle University\\
 %             Newcastle-upon-Tyne\\
 %             United Kingdom\\
 %             \email{carlton.shepherd@ncl.ac.uk}           %  \\


\maketitle

\begin{abstract}
Recent advances in generative artificial intelligence (AI), such as ChatGPT, Google Gemini, and other large language models (LLMs), pose significant challenges for maintaining academic integrity within higher education. This paper examines the structural susceptibility of a certified M.Sc.\ Cyber Security program at a UK Russell Group university to the misuse of LLMs. Building on and extending a recently proposed quantitative framework for estimating assessment-level exposure, we analyse all summative assessments on the program and derive both module-level and program-level exposure metrics. Our results show that the majority of modules exhibit high exposure to LLM misuse, driven largely by independent project- and report-based assessments, with the capstone dissertation module particularly vulnerable. We introduce a credit-weighted program exposure score and find that the program as a whole falls within a high to very high risk band. We also discuss contextual factors---such as block teaching and a predominantly international cohort---that may amplify incentives to misuse LLMs. In response, we outline a set of LLM-resistant assessment strategies, critically assess the limitations of detection-based approaches, and argue for a pedagogy-first approach that preserves academic standards while preparing students for the realities of professional cyber security practice.
\end{abstract}

\section{Introduction}

Large language models (LLMs) are a family of artificial intelligence (AI) systems that learn from very large collections of text to predict and generate language in context. Rather than being built for a single, narrow task, the same model can be prompted to summarise articles, explain concepts, translate between languages, or write computer code~\cite{kocmi2023large,kaddour2023challenges,ajevski2023chatgpt}. In practical terms, LLMs provide a general-purpose ``text partner'': users type a question or instruction in natural language and receive an apparently fluent response in return. Public-facing tools have dramatically lowered the barrier to access, making it straightforward for students and staff with no technical background to produce essays, reports, and technical artefacts on demand~\cite{kocmi2023large,kaddour2023challenges,ajevski2023chatgpt}. These tools are now widely used across sectors including finance, healthcare, law, and information technology~\cite{yang2024harnessing,kaddour2023challenges,nelson2024other}, and ChatGPT in particular ranks among the most visited websites globally~\cite{similarweb_top_websites}.

The development of LLMs is rooted in earlier generations of generative models. Initial approaches such as $n$-grams~\cite{damashek1995gauging} treated language as distributions over fixed-length token sequences. These were superseded by neural sequence models, notably recurrent neural networks (RNNs) and their variants, including long short-term memory (LSTM) and gated recurrent units (GRUs)~\cite{schmidhuber2015deep,goodfellow2016deep,dey2017gate}. Such models introduced mechanisms for capturing dependencies across variable-length contexts, but still struggled with long-range interactions and efficient training on very long sequences~\cite{goodfellow2016deep,attentionallyouneed}. The Transformer architecture of Vaswani et al.~\cite{attentionallyouneed} addressed many of these limitations through self-attention, enabling effective large-scale pre-training on diverse corpora. Modern LLMs build on this paradigm, using pre-training followed by fine-tuning or adaptation for specific tasks, and have demonstrated strong performance on translation, summarisation, and a variety of generative tasks, including creative writing and code generation~\cite{swanson2021story,kaddour2023challenges,kocmi2023large}.

The rapid popularisation of LLMs has had a marked impact on higher education. Some authors explicitly compare tools such as ChatGPT, Google's Gemini, and Anthropic’s Claude to the introduction of calculators in mathematics education~\cite{nelson2024other}. Usage data indicates that generative AI is already embedded in students' study practices. Von Garrel and Mayer~\cite{von2023artificial} report on a large-scale survey of German university students ($n=6311$) across 395 institutions: 63.4\% indicated that they had used AI-based tools for their studies. The most common use case was clarifying and explaining subject content (56.3\%). Other reported uses included developing research and literature studies (28.6\%), translations (26.6\%), text analysis and generation (24.8\%), and problem-solving and decision-making (22.1\%). When evaluating AI tools, students prioritised error prevention (e.g.\ mitigation of hallucinations\footnote{A \emph{hallucination} refers to instances where a generative artificial intelligence model produces content that is incorrect, fabricated, or nonsensical, despite appearing plausible or convincing.}), the perceived scientific quality of responses, and the tool's ability to reason. Table~\ref{tab:pref-vonmayer} summarises these preferences.

In this paper, we focus specifically on cyber security, which has received relatively little attention in the education literature with respect to LLM misuse. Cyber security professionals operate in environments where incompetence can lead to serious consequences, including data breaches, financial loss, and, in the most sensitive domains, threats to national security. If LLMs are exploited to obtain degrees without the genuine acquisition of knowledge and skills, unqualified individuals may enter the workforce under the guise of accredited competence. The integrity of assessments in cyber security programs is therefore central to ensuring that graduates possess the theoretical understanding and practical capabilities required for safe practice. Our analysis is motivated by this risk: we examine how the structure and composition of assessments in a certified M.Sc.\ cyber security program create opportunities for LLM misuse, and how this risk can be quantified at module and program level.

\begin{table}
\centering
\caption{Preferences when assessing generative AI tools by university students when asked \emph{``Which aspects are/were most important to you in your assessment?''} from~\cite{von2023artificial}.}
\resizebox{0.9\linewidth}{!}{%
\begin{tabular}{lrr}
\toprule
\textbf{Characteristics} & \textbf{$N$} & \textbf{\%} \\
\midrule
Error prevention during output (e.g., hallucination) & 3550 & 56.3 \\
Degree of scientificity (e.g., citation) & 4875 & 77.2 \\
Logical reasoning (e.g., answers are comprehensible) & 2942 & 46.6 \\
Explainability of the decision (e.g., white box vs.\ black box) & 2176 & 34.5 \\
Error detection and correction during input (e.g., grammar) & 1601 & 25.4 \\
Price & 2555 & 40.5 \\
\bottomrule
\end{tabular}
}
\label{tab:pref-vonmayer}
\end{table}

\subsection{Contributions}

This work presents a quantitative analysis of a certified UK cyber security degree program with respect to its susceptibility to LLM misuse. To this end, we apply and extend the framework by Hickey et al.~\cite{hickey2024threat} for evaluating assessments from a full complement of Master's degree modules. We develop a \emph{program exposure metric}, extending the work in \cite{hickey2024threat}, to quantitatively measure the susceptibility of a program as a whole by weighing modules and the level of LLM misuse risk associated with their individual assessments.  We discuss some of the challenges associated with program delivery, e.g.\ block teaching and international cohorts, which may create an environment in which misuse is incentivised. Lastly, we discuss potential solutions and their benefits and drawbacks of alternative assessment formats for addressing these challenges.

\subsection{Ethics}

This study received ethical approval by the author's institution on 20th April 2023. The results have been shared with the Faculty, the Degree Program Director, and the teaching staff group in order to address potential risks, such as inadvertently influencing student assessment outcomes or exacerbating the misuse of automated tools.

\subsection{Paper Structure}

The remainder of this paper is organised as follows. \S\ref{sec:background} reviews related work on the use and risks of LLMs in higher education, focussing on academic integrity and assessment design. \S\ref{sec:seced} then motivates cyber security as a domain of special concern, outlining the applied, high-stakes nature of the discipline and the role of specialist degree programs in preparing practitioners. \S\ref{sec:case-study} introduces the program that forms our case study, describing its structure and assessment portfolio. In this section, we present the methodology for measuring the exposure of assessments and modules to LLM misuse, before deriving a program-level exposure metric that accounts for credit weightings. We then present and interpret the results, showing where and how exposure to LLM misuse is concentrated. \S\ref{sec:solutions} discusses potential responses and recommendations, such as alternative assessment formats and the role and limitations of detection-based approaches. Finally, \S\ref{sec:conc} concludes the paper and identifies directions for future work.

\section{Background}
\label{sec:background}

Recent work has examined the implications of LLMs for higher education, highlighting both opportunities for enhanced learning and risks to academic integrity. In legal education, Nelson~\cite{nelson2024other} argues that the central challenge is not to prevent LLM misuse, but to make legal education \emph{``such a joyful, meaningful, and engaging enterprise that learners will want to do it themselves -- rather than let computers have all the fun.''} They note that LLMs are already capable of passing professional examinations, including US bar exams~\cite{ABAJournal2023ChatGPTBarExam} and the Solicitors Qualifying Examination in England and Wales~\cite{ajevski2023chatgpt}. Nelson explicitly invokes the calculator analogy from the 1980s, where initial fears of ``dumbing down'' ultimately gave way to an understanding of calculators as tools that can deepen conceptual learning. In this view, LLMs should be treated less as a threat to be policed and more as a catalyst for rethinking assessment and learning design.

Mortlock and Lucas~\cite{mortlock2024generative} examine the role of LLMs, specifically ChatGPT, in pharmacy education. They report that ChatGPT can achieve passing performance on third-year undergraduate medical exams and professional pharmacy board accreditation tests (see also Gilson et al.~\cite{gilson2023does}). Their analysis surfaces several concerns: (i) the \emph{``possible hindrance of essential skill development''}; (ii) the presence of factual inaccuracies in LLM outputs; (iii) inconsistencies between different LLMs; and (iv) inequities arising from differential access to paid services (e.g.\ premium tiers such as ChatGPT Pro, which offer \emph{``better answers to the hardest problems''}~\cite{openai_introducing_chatgpt_pro}). At the same time, they outline constructive uses of LLMs, including scenario generation, simulation of clinical encounters, and role-playing exercises in authentic learning environments.

Recent work has also aimed to foster more effective assessment designs. Chaudhary et al.~\cite{chaudhary2024exploring} proposed a framework for safely embedding generative AI into cyber security teaching that foregrounds transparency, critical scrutiny of AI outputs, and explicit discussion of security and privacy risks. Elkhodr and Gide~\cite{elkhodr2025integrating}, meanwhile, describe case studies in which students use generative AI to support policy analysis and risk assessment, but are required to critique and refine AI-generated content rather than accept it at face value. In computing education more generally, Zhang et al.~\cite{zhang2024students} show that students often perceive LLM-generated feedback on programming tasks as formative, and described the benefits of LLMs for improving feedback quality and staff efficiency.

A complementary strand of work has investigated the detectability of LLM-generated content in student submissions. Perkins et al.~\cite{perkins2024detection} evaluate the Turnitin AI detection tool using experimental submissions generated with ChatGPT that were crafted to evade detection. These were evaluated alongside genuine student work by 15 academic staff. Although the tool flagged 91\% of experimental submissions as containing some AI-generated content, only 54.8\% of that content was correctly distinguished in practice, and staff referred 54.5\% of flagged submissions to academic misconduct processes. Marking outcomes were similar between AI-generated and genuine work. The authors call for rethinking assessment strategies rather than relying primarily on detection, and they suggest reducing dependence on assignment formats that are highly susceptible to misuse.

Student attitudes and usage patterns have also begun to receive attention. Rogers et al.~\cite{rogers2024attitudes} report on an online survey of $n=70$ university-level computer science students regarding their use of ChatGPT and their ethical stance towards it. Almost all respondents (98.6\%) had heard of ChatGPT and 63.8\% had an active account. Students reported primarily using it as a search engine and for explaining unclear concepts from class. The authors conclude that students were \emph{``generally...using it to learn. There is, however, a small, but noteworthy minority of students that find little moral issue with submitting work produced by the tool.''} Notably, 65\% of students were undecided or disagreed that using ChatGPT to write assignments constitutes cheating, and only 28\% disagreed that doing so violates university ethical standards. A \emph{``notable minority''} openly admitted to using the tool in ways that could interfere with their learning~\cite{rogers2024attitudes}.

A methodological challenge in this area is obtaining honest self-reports about misuse. Student surveys, focus groups, and interviews are likely to be affected by social desirability bias: participants may underreport undesirable behaviours  or overreport desirable ones, even under assurances of confidentiality~\cite{zerbe1987socially}. This problem is well-documented in the wider literature~\cite{scheers1987improved,miller2011investigating,miller2008under} and complicates attempts to estimate the prevalence of LLM misuse from self-reported data alone. We take a different perspective. Rather than attempting to measure assessment design or students' preferences or use of LLMs, we analyse the structural susceptibility of an existing, externally certified M.Sc.\ cyber security program to LLM misuse. Building on the quantitative framework of Hickey et al.~\cite{hickey2024threat} for estimating assessment-level exposure to LLMs, we evaluate the extent to which the program's assessment portfolio, at a module and program level, creates opportunities for students to misuse LLMs. This allows us to contribute a discipline-specific data point for cyber security education and to propose a program-level exposure metric that can inform curriculum review and risk analysis.

\section{Security Education as an Area of Concern}
\label{sec:seced}

Cyber security is inherently a practice-oriented discipline in which theoretical understanding must be reinforced by hands-on experience. Academic programs typically incorporate practical components such as penetration testing labs, vulnerability assessment exercises, and incident response simulations~\cite{zeng2018improving,mirkovic2012teaching,vsvabensky2021cybersecurity}. These activities are intended to develop a deep understanding of attack vectors and defensive techniques, as well as the real-time decision-making skills required to respond to threats in authentic environments. In many curricula, such practical work is also central to assessment, through lab reports, practical write-ups, and project-based tasks that purport to evidence applied competence.

Cyber security education as a distinct academic domain is relatively recent compared with longer-established disciplines such as philosophy, mathematics, or even computer science. It is only since the 1990s that the field has gained a sustained foothold in mainstream higher education; in the case of cryptography, for example, the teaching of core techniques outside national security and intelligence contexts was effectively normalised only around 2000 (see \emph{Bernstein v.\ United States}~\cite{ross1998bernstein}). Early specialist modules and degree programs began to appear in the mid-1990s, and the field has often been likened to accounting and finance in terms of its applied character and the potentially consequential nature of errors~\cite{knapp2017maintaining}. Contemporary programs are frequently aligned with external accreditation or certification frameworks, which further emphasise demonstrable professional competence rather than purely theoretical mastery.

The area shares many of the challenges faced by computer science programs~\cite{rogers2024attitudes}, but these are amplified by the high-stakes context in which graduates operate. Students may be tempted to rely on LLMs to generate solutions or reports for practical assessments; for instance, scripting attacks (e.g.\ exploits), drafting penetration test reports, or describing vulnerabilities in simulated environments. Offloading substantial parts of the reasoning process to an LLM may short-circuit the development of the professional judgment that programs are designed to cultivate. Such misuse risks superficial or incorrect understanding of core concepts, underdeveloped investigative and analytical skills, and  bypassing formative experiences that are critical to professional competence.  Analogous concerns have been raised in other applied, high-responsibility fields such as medicine~\cite{mortlock2024generative,gilson2023does} and the law~\cite{nelson2024other}.

The pipeline from higher education into the cyber security workforce underscores the significance of these risks. A recent survey by the UK Department for Science, Innovation and Technology (DSIT)~\cite{dcms2024} reports that most frontline security operations and consultancy roles require a graduate degree; approximately 42\% of cyber firms employ staff with specialist higher education qualifications in cyber security, and around one quarter of staff hold a specialist IT or cyber security degree. A complementary US study of over 11{,}000 job postings in 2020 found that roughly 60\% of entry-level roles required a college degree, 24\% a graduate degree, and 29\% expressed a preference for specialist cyber security certification~\cite{marquardson2020skills}. In this context, the credibility of assessments is not merely an internal quality-assurance concern but a matter of workforce reliability. Employers and accrediting bodies implicitly trust that graduates who have passed programs possess the knowledge and competence to operate safely. Systematic misuse of LLMs to inflate assessment performance undermines this trust, obscuring the distinction between genuinely competent graduates and those who have leveraged automation to obtain credentials. If students progress through programs without authentically developing technical, analytical, and strategic competencies, they pose a tangible risk to the organisations and infrastructures they are later tasked with protecting. This makes cyber security a particularly important domain in which to understand how program structures may enable misuse and thus potentially posing a wider cyber security risk to society.

\section{Case Study: M.Sc. Cyber Security Program}
\label{sec:case-study}

Our work considers a M.Sc.\ Cyber Security at a Russell Group University within the United Kingdom. The degree has been certified by a nationally leading external authority with regards to the quality and rigour of its content. The program has an enrollment of approximately 20--40 students per annum of which $>$85\% are international students. Oftentimes, students are studying at the institution using English as the working language for the first time ($\approx$80\%). In the program, students must study a total of 180 credits across the academic year, split into 60 credits per semester as shown in Table~\ref{tab:course-modules}. The first semester is focussed on learning core competencies in cyber security fundamentals, such as key security notions (e.g.\ confidentiality, integrity, availability), risk, trust, and an introduction to cryptography and system security. Students may select one 10-credit module outside of these fundamental modules on Cloud Computing or Machine Learning.  The second semester involves more advanced modules, such as the security of complex systems, network security and ethical hacking, and a research methods module. Semester 2 also contains the program's only module taught in a traditional linear mode, i.e.\ lasting the whole semester. The final semester is dedicated to the completion of an independent research project culminating in the production of a dissertation and demonstration. 

The case study program is taught principally in block mode: a total of 8/10 modules (80\%) worth 110/160 credits ($\approx$70\%) are taught in this format. A small, 10-credit group-based advanced topics module and the dissertation project are the only exceptions to this delivery mode. This is not unusual within the university; the vast majority of postgraduate taught modules are delivered using block teaching within the department concerned.

\begin{table}
\centering
\renewcommand{\arraystretch}{1.5}
\caption{Case study modules. Rows give those in block delivery; M10 and M11 are delivered traditionally in Sem.\ 2 and 3 respectively. Credits in brackets.}
%\resizebox{\linewidth}{!}{%
\begin{tabular}{|p{3.9cm}|p{3.9cm}|p{0.3cm}|p{1.1cm}|}
\hline
\textbf{Sem.\ 1} & \multicolumn{2}{l|}{\textbf{Sem.\ 2}} & \textbf{Sem.\ 3} \\ \hline
\parbox[t]{3.9cm}{%
    [M1] Information Security \& Cryptography (10)\\\newline
    [M2] Risk \& Trust (10)\newline}
  &
\parbox[t]{3.5cm}{%
    [M7] Security of Complex Systems (10)}
  &
\multirow{10}{*}{%
      \centering
      \rotatebox{90}{%
          \shortstack{%
              [M10] Advanced Topics (10)%
          }%
      }%
    } 
  &
  
\multirow{10}{*}{%
      \rotatebox{90}{%
      \centering
          \shortstack{%
          \\\newline\\\newline
               [M11] Dissertation Project (60)%
          }%
      }%
    } \\ \cline{1-2}
\parbox[t]{3.5cm}{[M3] Systems Security (20)}
  & [M8] Research Methods and Group Project (20)
  &  & \\ \cline{1-2}
\parbox[t]{3.3cm}{%
    [M4] Secure Software Development (10)\\\newline
    [M5] Cloud Computing (10) \underline{\textbf{OR}} [M6] Machine Learning (10)\newline}
  & \parbox[t]{3.3cm}{[M9] Network Security \& Ethical Hacking (20)}  
  &  & \\ \hline
\end{tabular}
%}
\label{tab:course-modules}
\end{table}

\begin{table}
\centering
\resizebox{0.95\linewidth}{!}{%
\begin{threeparttable}
\caption{Case study module descriptions.}
\label{tab:modules}
\begin{tabular}{@{}p{0.1\linewidth}p{0.25\linewidth}p{0.1\linewidth}p{0.55\linewidth}@{}}
\toprule
\textbf{ID} & \textbf{Title} & \textbf{Credits} & \textbf{Description} \\ 
\midrule
\textbf{[M1]}\label{m1} 
  & Information Security \& Cryptography & 10 
  & Covers fundamental techniques including cryptography, PKI, and access control. Emphasises technical concepts and formal security methods. \\ 
\midrule

\textbf{[M2]}\label{m2} 
  & Risk \& Trust & 10 
  & Introduces systematic risk assessment, control measures, and security policies. Students analyse legal, ethical, and professional issues in security and trust management. \\ 
\midrule

\textbf{[M3]}\label{m3} 
  & Systems Security & 20 
  & Covers system vulnerabilities, threat modeling, and security engineering. Topics include cryptographic techniques, distributed system security, and cyber-physical systems. \\\midrule
  
\textbf{[M4]}\label{m4} 
  & Secure Software Development & 10 
  & Explores secure software engineering principles such as model-driven security and formal verification. Students learn to integrate secure practices into the software development lifecycle. \\ 
\midrule

\textbf{[M5]}\textsuperscript{\dag}\label{m5} 
  & Cloud Computing & 10 
  & Focuses on the principles of cloud computing, including architecture, virtualization, and security challenges. Topics include scalable computing patterns, virtual machines, and secure service design. Case studies provide practical insights into real-world applications. \\ 
\midrule

\textbf{[M6]}\textsuperscript{\dag}\label{m6} 
  & Machine Learning & 10 
  & Provides applied knowledge of machine learning, including classification, regression, and clustering. Students complete a project applying machine learning to an authentic problem. \\ 
\midrule

\textbf{[M7]}\label{m7} 
  & Security of Complex Systems & 10 
  & Addresses securing interconnected systems like IoT and smart grids. Topics include vulnerability analysis, human factors, and advanced attacks such as side-channel exploitation. \\ 
\midrule

\textbf{[M8]}\label{m8} 
  & Research Methods and Group Project & 20 
  & Students complete a group project applying security principles to real-world problems. Develops students' knowledge and skills in research methodologies, evaluation techniques, and collaborative project work. \\ 
\midrule

\textbf{[M9]}\label{m9} 
  & Network Security \& Ethical Hacking & 20 
  & Focuses on penetration testing, ethical hacking principles, and network security. Students gain hands-on experience with vulnerability assessments and security tools. \\ 
\midrule

\textbf{[M10]}\label{m10} 
  & Advanced Topics & 10 
  & Examines advanced trends like AI, Internet of Things, post-quantum security, and distributed ledger technology. The module emphasises critical evaluation of research and future trends. \\ 
\midrule

\textbf{[M11]}\label{m11} 
  & Dissertation Project & 60 
  & A capstone project involving independent research under supervision. Students develop a security-based artefact, e.g., application, algorithm, user study, culminating in a dissertation. \\ 
\bottomrule
\end{tabular}
\begin{tablenotes}
\item[\dag] In Semester 1, students select either [M5] \underline{\textbf{or}} [M6] as an elective module, not both.
\end{tablenotes}
\end{threeparttable}
}
\end{table}

%\subsection{Delivery Format}

\subsection{Assessment Breakdown}

The M.Sc.\ Cyber Security program is structured to develop students' theoretical knowledge and practical skills. This is assessed through a variety of means, from written essays, traditional examinations, oral examinations, lab reports, group reports, online quizzes, and presentations. The assessments vary in type, complexity and weighting. Note that there is no strict limit on the number and type of assessments that may be used, e.g.\ at a School level; however, each module has at least one graded summative assessment. Many modules also contain formative assessments, which have nil contribution to students' overall grades, but which are designed to support their learning. Our work concentrates on the summative assessments that contribute to their overall degree outcomes. We briefly describe the assessments for each module and their types in Table~\ref{tab:modules}.

\subsection{How Risky Are The Assessments?}

We adopt the framework of Hickey et al.~\cite{hickey2024threat},  which estimates the potential exposure of higher education assessments to LLM misuse and is intended to support the \emph{``planning and reviewing teaching and learning practices and policies''}. An assessment is judged to be exposed to LLM misuse based on the following:
\begin{itemize}
    \item The \textbf{assessment type}, e.g.\ an oral examination is deemed to have lower exposure than a take-away essay.
    \item The \textbf{contribution} of the assessment to the module grade. 
    \item The \textbf{intended location} in which the assessment occurs. A practical examination in the presence of invigilators, for instance, is considered to have lower exposure than an assignment that is completed at home by the student where no supervision or invigilation is in place.
\end{itemize}

\begin{table}
\centering
\caption{Scoring LLM-exposed assessments from~\cite{hickey2024threat}.}
\label{table:llm_exposure}
\begin{tabular}{rccc}
\toprule
\textbf{Assessment Type} & \textbf{Location} & \textbf{Indicator} & \textbf{Score} \\ 
\midrule

% ----------------------- In Person -----------------------
Attendance          & \multirow{9}{*}{In Person} & \cellcolor{green!70}  & 0 \\
Class Test          &                             & \cellcolor{green!70}  & 0 \\
Examination         &                             & \cellcolor{green!70}  & 0 \\
Fieldwork           &                             & \cellcolor{green!70}  & 0 \\
Oral Examination    &                             & \cellcolor{green!70}  & 0 \\
Practical Examination &                           & \cellcolor{green!70}  & 0 \\
Presentation        &                             & \cellcolor{green!70}  & 0 \\
Seminar             &                             & \cellcolor{green!70}  & 0 \\
Studio Examination  &                             & \cellcolor{green!70}  & 0 \\
\midrule

% ----------------------- Blended -------------------------
Lab Report                   & \multirow{5}{*}{Blended} & \cellcolor{yellow!70} & 2 \\
Multiple Choice Questionnaire&                          & \cellcolor{yellow!70} & 2 \\
Portfolio                    &                          & \cellcolor{yellow!70} & 2.5 \\
Group Project                &                          & \cellcolor{orange!70} & 3 \\
Journal                      &                          & \cellcolor{orange!70} & 3 \\
\midrule

% ----------------------- At Home -------------------------
Assignment           & \multirow{4}{*}{At Home} & \cellcolor{red!70} & 5 \\
Continuous Assessment&                         & \cellcolor{red!70} & 5 \\
Essay                &                         & \cellcolor{red!70} & 5 \\
Project              &                         & \cellcolor{red!70} & 5 \\

\bottomrule
\end{tabular}
\end{table}

Hickey et al.~\cite{hickey2024threat} categorise the risk of various assessment types, given in Table~\ref{table:llm_exposure}. Their work gives a module \emph{exposure estimate} metric for estimating the vulnerability of assessments:

\begin{equation}
  \mathrm{ModuleExposure} = \sum_{n=1}^{N} R_n\,S_n,\qquad S_n\in[0,1],\ \sum_{n} S_n=1.
  \label{eq:exposure-estimate}
\end{equation}

Where $R_n$ is the exposure score for a given assessment based on Table~\ref{table:llm_exposure}, $S_n$ is the contribution of that assessment to the final module grade as a percentage, and $N$ is the number of assessments for a given module.

Other informal measures have been suggested for evaluating assessments~\cite{csu_rethinking_assessment_2023,perkins2024detection,mortlock2024generative}. However, we adopt a quantitative approach to help easily compare modules with respect to their exposure to LLM misuse. The aforementioned exposure metric (Eq.~\ref{eq:exposure-estimate}) yields scores in the range $[0,5]$, where 0 represents low risk and 5 the highest risk to misuse.  We applied this to all of the module assessments in the program case study, mapping the assessments to their type based on the module outline forms. We then give LLM Exposure Scores (LLMESs) on two bases: firstly, on the basis of individual assessments, from Table~\ref{table:llm_exposure}; and, secondly, the weighted LLMES for each module using Eq.~\ref{eq:exposure-estimate}. Table~\ref{tab:assessments_llm} presents the data following the application of this procedure. Histograms for both individual assessment- and module-level scores are given in Figure~\ref{fig:llmes-comparison}.

\subsection{Findings}

\begin{figure}[t]
    \centering
    % First Subfigure
    \begin{subfigure}{0.44\linewidth}
        \centering
        \includegraphics[width=\linewidth]{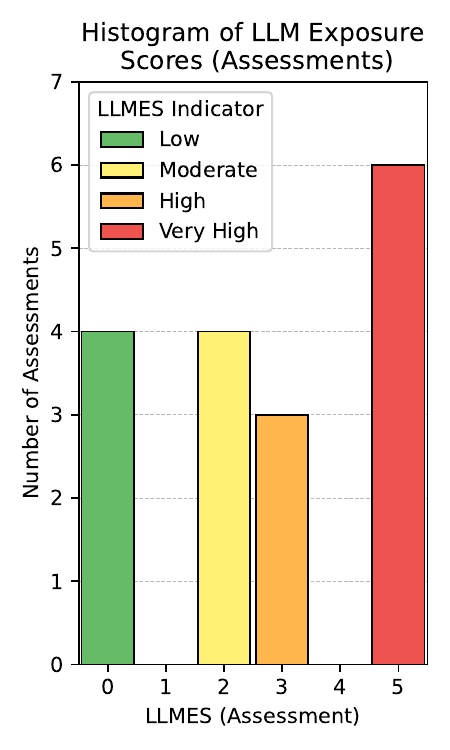}
        \caption{}
        \label{fig:llmes-histogram}
    \end{subfigure}
    % Second Subfigure
    \begin{subfigure}{0.44\linewidth}
        \centering
        \includegraphics[width=\linewidth]{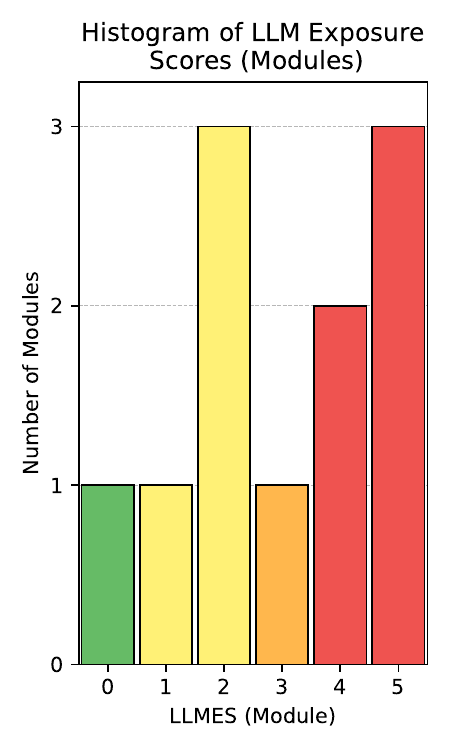}
        \caption{}
        \label{fig:llmes-weighted}
    \end{subfigure}
    % Main Caption
    \caption{Distributions of LLM exposure scores (LLMESs) for individual assessments~(\subref{fig:llmes-histogram}) and on a module basis~(\subref{fig:llmes-weighted}).}
    \label{fig:llmes-comparison}\vspace{-0.5cm}
\end{figure}

The results indicate that the program's susceptibility to LLM misuse is substantial. Across the 17 summative assessments identified in Table~\ref{tab:assessments_llm}, LLMESs span the full range from 0 (low exposure) to 5 (very high exposure). At the assessment level, 6/17 (35.3\%) assessments are assigned the maximum score of 5, and 9/17 (52.9\%) have scores of 3 or above, indicating medium-to-very-high exposure. Only 4/17 (23.5\%) assessments fall into the lowest exposure band with a score of 0. The distribution in Fig.~\ref{fig:llmes-histogram} is therefore skewed towards the upper end of the scale, with a concentration of assessments in the higher-risk categories.

\newcolumntype{P}[1]{>{\centering\arraybackslash}p{#1}}
\newcolumntype{R}[1]{>{\raggedleft\arraybackslash}p{#1}}
\begin{table}
\centering
\renewcommand{\arraystretch}{1.5}
\caption{Assessment LLM Exposure Scores (LLMES).}
\label{tab:assessments_llm}
\resizebox{\linewidth}{!}{%
\begin{tabular}{@{}R{1cm}|l|p{5.5cm}|c|P{1.5cm}|P{1.25cm}@{}}
\toprule
\textbf{ID} & \textbf{Type} & \textbf{Assignment Description} & \textbf{\%} & \textbf{LLMES (Assessment)} & \textbf{LLMES (Module)} \\ \midrule
\textbf{[M1]} & Written Exam & Traditional examination covering relevant topics on information security and cryptography. & 100 & \cellcolor{green!70} 0 & \cellcolor{green!70} 0 \\ \midrule
\textbf{[M2]} & Report & 1000 words, plus listings, references, and figures. & 100 & \cellcolor{red!70} 5 & \cellcolor{red!70} 5 \\ \midrule
\textbf{[M3]} & Quiz & Canvas online quizzes on system security. & 40 & \cellcolor{yellow!70} 2 & \cellcolor{red!70} 3.8 \\ 
 & Report & 1500 words with listings and documentation. (10 pages) & 60 & \cellcolor{red!70} 5 & \cellcolor{red!70} \\ \midrule
\textbf{[M4]} & Lab Report & Exercises in secure programming, logic, and verification. & 100 & \cellcolor{yellow!70} 2 & \cellcolor{yellow!70} 2 \\ \midrule
\textbf{[M5]} & Lab Report & Testing of programming skills; code demonstration + report. & 40 & \cellcolor{yellow!70} 2 & \cellcolor{yellow!70} 0.8 \\ 
 & Exam & Invigilated, computer-based examination. & 60 & \cellcolor{green!70} 0 &  \cellcolor{yellow!70} \\ \midrule
\textbf{[M6]} & Project & Machine learning project implementation and report. & 100 & \cellcolor{red!70} 5 & \cellcolor{red!70} 5 \\ \midrule
\textbf{[M7]} & Assignment & A 2000-word assignment on complex systems security. & 100 & \cellcolor{red!70} 5 & \cellcolor{red!70} 5 \\ \midrule
\textbf{[M8]} & Group Project & Group literature review (20 pages) & 30 & \cellcolor{orange!70} 3 & \cellcolor{orange!70} 3 \\ 
 & Group Project & End of module, group report (13 pages) & 70 & \cellcolor{orange!70} 3 &\cellcolor{orange!70} \\ \midrule
\textbf{[M9]} & Lab Report & A security analysis relating to individual exercises completed in the practical sessions. & 100 & \cellcolor{yellow!70} 2 & \cellcolor{yellow!70} 2 \\ \midrule
\textbf{[M10]} & Presentation & Group presentation where all group members must take part. & 25 & \cellcolor{green!70} 0 & \cellcolor{yellow!70} 2.25 \\ 
 & Group Project & Group report on an advanced cyber security topic with an individual reflection. & 75 & \cellcolor{orange!70} 3 & \cellcolor{yellow!70} \\ \midrule
\textbf{[M11]} & Assignment & Interim report, with aim, objectives, and plan. & 5 & \cellcolor{red!70} 5 & \cellcolor{red!70} \\ 
 & Oral Exam & 30 min.\ viva (20 min.\ presentation + 10 min.\ Q\&A). & 20 & \cellcolor{green!70} 0 & \cellcolor{red!70} 4 \\ 
 & Project & Independent research project on cyber security. & 75 & \cellcolor{red!70} 5 & \cellcolor{red!70} \\ \bottomrule
\end{tabular}
}
\end{table}

When these assessments are aggregated to the module level, a similar pattern emerges. Module LLMESs range from 0 to 5, with median of $\approx$3, placing the typical module in the high-exposure band. Only a single module, [M1], is classified as low risk, based on a traditional invigilated written examination. Five modules have low-to-moderate exposure ($<$3), owing to the use of lab reports or blended assessments (e.g.\ [M4], [M5], [M9], [M10]). By contrast, six modules fall into the high or very high bands ($\geq$3), including [M2], [M3], [M6], [M7], [M8], and [M11]. The module-level histogram in Fig.~\ref{fig:llmes-weighted} shows a clear shift towards higher exposure once assessment weights are taken into account. We observe generally that the heavy use of take-home, text-based assessments by modules ([M2], [M3], [M6], [M7], and [M11]) are a major contributing factor to high exposure scores---aligning closely with the strengths of current LLMs. By contrast, assessments of moderate exposure are predominantly lab reports and online quizzes linked to practical work ([M3], [M4], [M5], [M9]), while those group reports and literature-based projects have higher misuse exposure ([M8], [M10]). The four lowest-exposure assessments are all invigilated or live-performance formats: two examinations, a presentation, and an oral viva ([M1], [M5], [M10], [M11]). %Taken together, this confirms that the case-study program relies heavily on assessment types that current frameworks regard as especially vulnerable to LLM misuse. 

The structure of exposure across the curriculum is also informative. In Sem.~1, only [M1] is low-risk; other core modules include high-exposure elements such as a 100\% report in [M2], a combined quiz and report structure in [M3], and a 100\% project in [M6]. Sem.~2 retains this pattern: [M7] and [M8] both rely heavily on at-home assignments and group reports, while [M9] and [M10] sit in the moderate-exposure range due to their mixture of lab-based work and group projects. Sem.~3 is dominated by the dissertation module ([M11]), which, with a module LLMES of 4.0, is among the most exposed components of the program. In aggregate, this means that students encounter high-exposure assessments throughout the entire academic year rather than only in isolated modules. 

We also observe that low-exposure components can be diluted at the module level by higher-risk elements. For example, [M10] combines a low-risk presentation (25\%) with a higher-risk group report (75\%), yielding an overall module LLMES of 2.25. Similarly, [M11] includes an oral examination (20\%) and a take-home project and interim report (80\%), resulting in a high module score (4.0) despite the presence of a viva. Structurally, this means that adding an oral or invigilated component does not necessarily shift a module out of the high-risk band if the dominant contribution to the grade remains unsupervised written submissions. The assessment portfolio generally shows a disproportionate reliance on take-home, text-centric assignments and projects. This creates multiple points in the program where students could, in principle, substitute LLM-generated artefacts for their own work with relatively low likelihood of immediate detection. \S\ref{sec:solutions} considers how such structural patterns might be adjusted by rebalancing assessment tasks to reduce the value of straightforward LLM substitution.

\subsection{A Program-level Perspective}

The previous results focussed on assessment- and module-level exposure scores. We may also wish to evaluate the susceptibility of a \emph{program} at large to LLM misuse. To this end, we extend the scoring metric in Eq.~\ref{eq:exposure-estimate} from \cite{hickey2024threat} to weight modules' assessment contributions to the program as a whole:

\begin{equation}
  \mathrm{ProgramExposure} = \sum_{i=1}^{M} \frac{\mathrm{Credits}(i)}{\mathrm{TotalCredits}}
  \sum_{j=1}^{N_i} R_{i,j}\,S_{i,j}.
\end{equation}

In our case study, `Total Credits' equals 180 in line with the standard program credits for a postgraduate UK degree. `Credits($i$)' represents the number of credits for the $i^{th}$ module on the program. This is a useful value to consider because the effects of high-risk modules may be reduced if their contribution to the overall program is smaller on aggregate. We computed the results for the case study using this equation. This yielded Program Exposure values of \underline{\textbf{3.10}} and \underline{\textbf{3.34}}, depending on whether the student takes [M5] or [M6], respectively, as an optional module. As such, we can  characterise the program as having `High' to `Very High' exposure to LLM misuse if one extends the LLMES indicators to a program level (see Table~\ref{table:llm_exposure} and Fig.~\ref{fig:llmes-comparison} for bandings).  The decomposition of the program exposure reveals how risk is distributed across modules. Unsurprisingly, the 60-credit [M11] (Dissertation) is the single largest contributor: with a module LLMES of 4.0 and a one-third share of the total credits, it accounts for more than 33\% of the overall program exposure. High-exposure 10- and 20-credit modules such as [M2], [M3], [M6], [M7], and [M8] then add further increments, while lower-exposure modules such as [M1], [M4], [M5], [M9], and [M10] have only limited moderating effect. 

We note that this formulation also illustrates how the metric can support ``what-if'' reasoning for curriculum design. For example, replacing a high-exposure assessment in a 10-credit module with a low-exposure, invigilated alternative reduces the program score only marginally, whereas altering the structure of the 60-credit dissertation module would have a much larger impact. In addition, substituting a moderate-risk elective ([M5]) for a high-risk one ([M6]) increases the program exposure by roughly 0.2–0.3 points, making explicit that individual students' pathways can meaningfully alter their aggregate exposure to LLM-susceptible assessment conditions. The qualitative picture remains the same irrespective of the elective module choice: in both paths, the program-level exposure is dominated by large, highly exposed modules. 

It is important to highlight that our extension remains structurally aligned with~\cite{hickey2024threat}, inheriting their focus on assessment type, weighting, and location, but aggregates these quantities to the program level via credit weights. It does \emph{not} incorporate broader contextual factors that may modulate actual misuse behaviour. For instance, it is agnostic to cohort composition (e.g.\ proportions of home vs.\ international students), students' prior experience of UK higher education, language proficiency, and delivery modes. It also ignores institutional support structures such as academic integrity training, access to writing support, and pastoral provision, all of which may influence when and why students turn to LLMs. The values reported above should thus be interpreted as conservative structural estimates. It is plausible that the program's \emph{actual} exposure to LLM misuse is higher once these contextual variables are considered. A fuller understanding of how these factors interact with structural exposure would require complementary empirical work (e.g.\ longitudinal surveys, misconduct case analysis, or learning analytics), which we defer to future research.

%It is important to highlight that Hickey et al.~\cite{hickey2024threat} do not account for several broader contextual factors in their evaluation of vulnerable assessment types. For instance, their study does not consider cohort composition---such as the proportion of home versus international students---nor does it take into account students' native languages, prior experiences in higher education, module delivery formats, or the impact of deadline scheduling. Moreover, the analysis overlooks the role of institutional support structures, including academic integrity training, access to learning resources, and student well-being and pastoral services, all of which could influence instances of misuse. Consequently, it is plausible to hypothesise that the program's \emph{actual} exposure to LLM misuse may be significantly higher when these additional variables are considered. However, further research is needed to fully understand the extent to which these factors influence students' propensity to (mis)use LLM tools.

\section{Potential Solutions and Recommendations}
\label{sec:solutions}

\subsection{Alternative Assessment Formats}

One response to LLM misuse is to adjust assessment formats towards those deemed to be at lower structural risk. This may involve the greater use of invigilated or live-performance assessments, such as written examinations, oral examinations, and in-class tests. Such formats can be delivered in controlled environments where assessors can more confidently attribute the work to the student~\cite{csu_rethinking_assessment_2023,hickey2024threat}. In the case study, informal discussions suggested that some modules were already considering a shift towards closed-book or in-lab assessments. However, simply ``doing more exams'' is not a panacea; short, high-stakes examinations are known to negatively impact student and instructor well-being and can incentivise surface learning~\cite{szabo2018exams,gharibyan2005assessing}. Moreover, this also risks misalignment with learning outcomes, particularly in practice-oriented domains such as cyber security where project work and authentic artefact development are desirable. Program-level reforms must balance reductions in LLM exposure against the costs to learning, fairness, and staff and student workload.

The delivery model observed in the case-study program offers further complications. The use of condensed `block' teaching, in which modules are delivered over 1--4 week periods, has been associated with increased fatigue, reduced opportunities for spaced practice, and challenges for active learning compared with traditional semester-long formats~\cite{sewagegn2021modular,burton2008block,daniel2000review}. Under such conditions, students face a high density of teaching and assessment activity in short time windows. It is reasonable to hypothesise that this creates additional pressure that incentivises the misuse of LLMs to produce reports or code under time constraints. The predominance of international students in the cohort, many of whom are working in English in a UK higher education context for the first time, may further amplify this pressure by adding linguistic and cultural load to already intensive study patterns. While our quantitative metric does not directly capture these factors, they point to an interaction between structural exposure (e.g.\ assessment formats) and contextual stressors that warrants further investigation.

Beyond rebalancing assessment formats, staff may wish to redesign existing coursework. One strand of work involves designing tasks that require students to demonstrate process as well as product; for example, submissions linked to in-class activities, or lab-based checkpoints where students reproduce or extend work completed independently. Another approach is to emphasise authentic, performance-based assessment that mirrors the complexity of professional practice. Examples include simulations, case-based analysis, and practical skill evaluations in which students must justify decisions, explain trade-offs, or interpret evidence. The cyber security education literature offers some concrete examples. Capture-the-flag (CTF) events~\cite{leune2017using}, cyber wargames~\cite{haggman2019wargaming}, and red/blue team exercises~\cite{workman2021study} immerse students in realistic, time-bounded scenarios that require hands-on investigative work and real-time decision-making. When assessed appropriately (e.g.\ through debriefs, logs, or oral explanations of strategy), such activities make it more difficult for an LLM to act as a replacement for student effort. That said, these assessments are not without cost: they are comparatively complex to design and maintain, do not always scale well to large cohorts, and often require dedicated technical infrastructure and expertise. We argue that a pragmatic strategy is likely to involve selective use of authentic formats at key points in the program, complemented by more modest redesign of coursework that moves away from polished, text-based end products.

\subsection{Detection Methods}

Another class of mitigation techniques is on detecting LLM-generated text in student submissions. Commercial and open-source tools, such as Turnitin's AI detection features, claim to infer authorship by analysing linguistic artefacts. While attractive in principle, the reliability of such systems is limited. Recent studies show that detection algorithms can exhibit substantial rates of false positives and negatives, and that their performance degrades as LLMs evolve or as students adopt more sophisticated evasion strategies~\cite{wang2023student,bellini2024between,dalalah2023false}. As such, detection tools are better viewed as weak signals for further human scrutiny than as robust evidence for supporting high-stakes academic decisions.

Beyond generic detectors, some instructors have begun experimenting with embedding ``LLM traps'' into assessments: for example, prompt injections or hidden text intended to elicit characteristic errors if a student copies the task verbatim into a generation tool~\cite{yu2023assessing,abril2023whitefont}. While such techniques may surface certain forms of misuse, they raise significant ethical and pedagogical concerns. They constitute a form of \emph{deceptive assessment}~\cite{eaton2023ai}, risk penalising students who are acting in good faith but whose work happens to trigger a trap, and implicitly reframe the assessment relationship as adversarial. Relying heavily on detection---automated, trap-based, or both---risks undermining trust more broadly within the learning environment~\cite{gorichanaz2023accused}. Students may fear being incorrectly accused, particularly those whose writing style deviates from expected or native norms. This concern is acute in diverse international cohorts, where linguistic variation itself may be misinterpreted as evidence of AI assistance. Here, punitive detection regimes can introduce inequities, discourage legitimate use of LLMs for permitted purposes (e.g.\ proofreading, brainstorming, translation), and damage students' confidence in the fairness of institutional processes.

These limitations suggest that detection methods should be used cautiously and within a broader strategy. Where detection tools are employed, institutions should be clear with students about how they are used and the evidential weight they carry, and provide clear routes for students to contest or explain flagged work. More fundamentally, however, the primary line of defence against LLM misuse is likely to lie in assessment and curriculum design by constructing modules and delivery models that reduce the payoff from illegitimate automation.%, while still allowing students to use LLMs in structured, explicit ways that support rather than replace learning.

\subsection{Constructive and Transparent Use of LLMs}

A third, complementary strand of response is to integrate LLMs into cyber security education in explicitly sanctioned ways. Rather than treating LLMs solely as a threat, programs can position them as tools that students are expected to use critically and within clear boundaries. As recent work suggests~\cite{chaudhary2024exploring,elkhodr2025integrating,zhang2024students} (see \S\ref{sec:background}), this may ultimately entail shifting from a purely prohibition-detection model towards one in which LLM use is designed into learning activities and made an object of reflection, rather than a hidden practice. In the context of the case-study program, this could involve, for example:
\begin{itemize}
  \item Designing tasks where the learning outcome is to evaluate LLM outputs; for instance, analysing vulnerabilities, flawed configurations, or weak security policies proposed by an LLM, and then correcting or hardening them.
  \item Using LLMs to generate realistic practice-based artefacts---e.g.\ synthetic phishing scenarios, log excerpts, consulting/incident narratives, or basic threat models---that students must respond to.
  \item Evaluating the security and privacy implications of LLMs (e.g.\ prompt injection and data leakage) in module learning outcomes, so that students treat them as security-relevant systems in their own right.
\end{itemize}

Of course, such approaches do not eliminate the risk of misuse, but they make LLM use visible, discussable, and assessable, and they align more naturally with the mindset that cyber security education aims to foster. However, it is arguably a substantial change in practice for academic staff, requiring significant investments to redesign assessments, develop sufficient AI literacy, and supervise and critique student use. Without corresponding institutional support, there is a danger that calls for `innovative' AI use become an additional unfunded burden rather than a sustainable shift in pedagogy.

\subsection{Limitations}

The findings in this study should be read with several caveats. Reducing a complex assessment portfolio to a single program-level exposure score is deliberately reductive; we posit that the metric is useful for comparison and for communicating risk to program leaders at a high level. It does not account for actual pedagogical quality or student outcomes, which may ameliorate the risks of misuse even in modules with high exposure scores. Further, the program-level metric assumes that a module's share of credits reflects its contribution to overall learning, which is not always the case: low-credit modules may play critical formative roles. The scoring also implicitly treats LLM capabilities and assessment types as static. In reality, both are changing: advances in LLMs and student practices may undermine current assumptions about the relative safety of particular formats, meaning that periodic recalibration will be required. Finally, our analysis is based on formal module documentation for a single UK program. We did not observe assessment enactment or student behaviour, nor did we model local institutional support or detection practices. As such, the numerical values reported should be interpreted as a context-specific snapshot and as a demonstration of method, rather than as definitive judgements about this or any other program.

\section{Conclusion}
\label{sec:conc}
This study set out to evaluate the susceptibility of a certified UK Master's program in cyber security to misuse of LLMs, with the dual aim of providing a characterisation of assessment-level risk and introducing a quantitative approach for examining risk at a program scale. Across all summative assessments, we found that most modules exhibit high exposure to LLM misuse, driven primarily by the prevalence of take-home reports, independent projects, and other unsupervised assessment formats. Only one module in the program was classified as low-risk, while several---including the dissertation---contained assessment structures almost entirely composed of high-exposure components. We also showed that the program as a whole falls within the `High' to `Very High' exposure bands (3.10–3.34), depending on elective selection, using a novel program-level exposure metric. This provides a more holistic understanding of susceptibility than assessment- or module-level analyses alone, and offers a generalisable methodology for institutions seeking to benchmark or audit their own programs.

These findings have important implications for cyber security education, where graduates are expected to demonstrate competence in high-stakes, practice-oriented domains. We note that the combination of block teaching, heavy reliance on written coursework, and a predominantly international cohort may amplify the incentive to misuse LLMs, underscoring the need for assessment formats that are both pedagogically robust and resistant to automation. While detection tools and adversarial assessment design are possible mitigations, they carry limitations and ethical concerns. As such, we argue that a greater emphasis on authentic, performance-based assessments may offer a more sustainable path forward. Lastly, in future work, we believe that an investigation in contextual factors---such as delivery mode, student workload, language proficiency, and institutional support---offers an interesting frontier for exploring misuse behaviours among learners in higher education. %Our program-level metric provides a foundation for such studies and can be adapted to evaluate the resilience of curricula as LLM capabilities evolve.

%\section*{Declaration}

%\subsection*{

%\subsection*{Data Availability}

%The data supporting the findings of this study are available from the author upon reasonable request.

%Carlton Shepherd thanks the Faculty of Science, Agriculture and Engineering (SAgE) at Newcastle University for funding this work though an internal education grant. He also thanks Essam Ghadafi and other members of the Secure and Resilient Systems (SRS) group within the School of Computing for insightful discussions on the topic.

\bibliographystyle{ieeetr}
\bibliography{refs}

%\section*{Author Biography}

%Dr.\ Carlton Shepherd is a Lecturer ($\sim$Assistant Professor) in Computer Science at Newcastle University, United Kingdom. Dr.\ Shepherd is a Fellow (FHEA) of the UK Higher Education Academy and a member of the EPSRC Peer Review College. He is an experienced module leader in the cyber security education, with research interests in system and network security. He is a co-investigator of the NCSC/GCHQ Academic Centre of Excellence in Cyber Security Research, and ethics lead for the School of Computing. Dr.\ Shepherd holds a PhD in Information Security from Royal Holloway, University of London, and a B.Sc.\ in Computer Science from Newcastle University. 

\end{document}